\documentclass[fleqn,10pt]{SelfArx} 

\usepackage[english]{babel} 

\definecolor{color1}{RGB}{0,0,90} 
\definecolor{color2}{RGB}{0,20,20} 

\usepackage{hyperref} 

\hypersetup{
	hidelinks,
	colorlinks,
	breaklinks=true,
	urlcolor=color2,
	citecolor=color1,
	linkcolor=color1,
	bookmarksopen=false,
	pdftitle={Title},
	pdfauthor={Author},
}

\JournalInfo{March} 
\Archive{2022} 

\PaperTitle{Exabel's Factor Model} 

\Authors{Øyvind Grotmol\textsuperscript{1}*, Michael Scheuerer\textsuperscript{2}, Kjersti Aas\textsuperscript{2}, Martin Jullum\textsuperscript{2}} 
\affiliation{\textsuperscript{1}\textit{Exabel, Oslo, Norway}} 
\affiliation{\textsuperscript{2}\textit{Department for Statistical modeling, Machine learning and Artificial intelligence, Norsk Regnesentral, Oslo, Norway}} 
\affiliation{*\textbf{Corresponding author}: grotmol@exabel.com} 

\newcommand{\keywordname}{Keywords} 

\Abstract{Factor models have become a common and valued tool for understanding the risks associated with an investing strategy. In this report we describe Exabel's factor model, we quantify the fraction of the variability of the returns explained by the different factors, and we show some examples of annual returns of portfolios with different factor exposure.}

\begin{document}

\maketitle 

\tableofcontents 

\thispagestyle{empty} 


\section*{Introduction} 

\addcontentsline{toc}{section}{Introduction} 

In the context of factor investing, `factors' are thought of as characteristics shared by a group of securities and acting as the drivers of their returns. While macroeconomic factors such as interest rates, economic growth rates, etc.\ capture broad risks across asset classes, the model considered here is based on fundamental factors such as each company's country and industry membership and financial characteristics (size, book value, earnings, etc.), which help explain returns and risks within asset classes.

Factors can be used to construct indices with increased exposure to a particular factor. Some factor indices have been demonstrated to earn persistent premiums over long time periods, but they often exhibit substantial cyclicality over short time horizons and may underperform for periods up to several years. Within a factor model, the main role of the factors is to help construct a low dimensional, interpretable model for the co-movement of a large number of asset returns. The risk associated with this set of returns -- and of portfolios based on it -- can then be decomposed into a risk component explained by the respective factor exposures and an idiosyncratic risk. The latter is specific to each security and assumed independent from the factor-based risks and across different securities. 

In this document we describe the details of Exabel's factor model. Further, we provide some figures which give an idea about how much of the variability of returns across all considered companies is explained by this factor model, and we finally calculate the annual returns of a few portfolios with increased exposure to a particular factor group to illustrate how these portfolios would have performed over the last ten years. Results of a more in-depth analysis of the factor model's ability to quantify the risks associated with different portfolios is provided in a separate document entitled `Performance evaluation of volatility estimation methods for Exabel'.


\section{Definition of the factor model}

\begin{figure*}[ht]\centering
	\includegraphics[width=\linewidth]{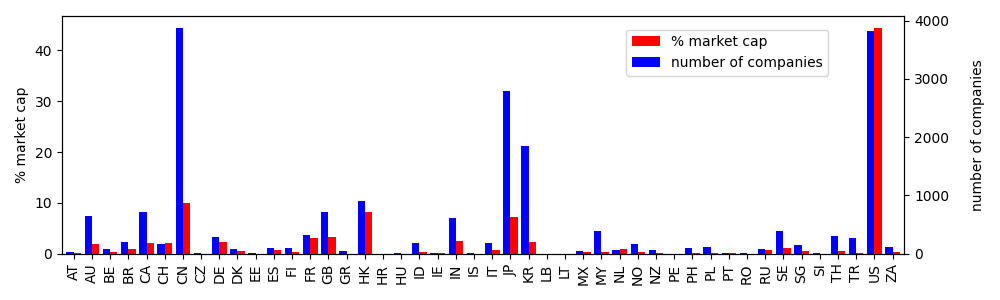}
	\caption{Number of companies in each country and associated share of the total market capitalization within the set of countries in the estimation universe for which all factor loading data was available on 1.1.2021.}
	\label{fig:country-statistics}
\end{figure*}

Denote by $r_k$ the return of a company $k\in\mathbb{K}:=\{1,\ldots,K\}$ on a given day. We call this set $\mathbb{K}$ of investible companies the {\em estimation universe}. At the time of writing, Exabel's estimation universe contains $K=28~629$ companies from all over the world\footnote{Note, however, that for some of these companies data might not be available for the entire evaluation period, so on a particular day the effective estimation universe $\mathbb{K}_{\mathrm{eff}}$ used in the cross section model can be much smaller.}.

With a factor model we would like to explain the returns of all companies in $\mathbb{K}$ as well as possible through a small number of interpretable factors, i.e.\ attributes of a company which appear to influence its return. The {\em factor loading} $X_k^{\ast}$ of company $k$ quantifies its exposure to a particular factor. Three different types of factor loadings are considered in our model:
\begin{itemize}[noitemsep] 
    \item {\em Style}-loadings $X_k^s, \; s\in\mathbb{S}$ (set of style factors)
    \item {\em Country}-loadings $X_k^c, \; c\in\mathbb{C}$ (set of countries)
    \item {\em Industry}-loadings $X_k^i, \; i\in\mathbb{I}$ (set of industries)
\end{itemize}
Using a cross section regression model, we then want to estimate (separately for each day) {\em factor returns} $f_s, s\in\mathbb{S}, f_c, c\in\mathbb{C}$, and $f_i, i\in\mathbb{I}$ together with the overall market return $f_m$:
\begin{equation}
 r_k = f_m + \sum_{s\in\mathbb{S}} X_k^s f_s + \sum_{c\in\mathbb{C}} X_k^c f_c + \sum_{i\in\mathbb{I}} X_k^i f_i + \epsilon_k, \label{eq:factor-model}
\end{equation}
where $\epsilon_k\sim\mathcal{N}(0,\sigma_k^2)$ is the residual term that accounts for the fraction of the return $r_k$ not explained by the company's exposure to the different factors.

\subsection*{Style factors}

The {\em style} loadings quantify characteristics of a company such as quality, growth, or value which have the potential to explain why portfolios with above or below average exposure to these factors have outperformed others. A wealth of different factors has been proposed in the literature over the last decades, but in order to ensure good interpretability of Exabel's factor model, it focuses on 11 style factors which were found to be particularly useful in explaining the differences in returns across the estimation universe. Table~\ref{tab:style-factors} lists these factors, organised into 8 factor groups proposed by MSCI \cite{MSCI-FaCS} to give an even more intuitive overview. Details and exact mathematical definitions of these factors are given in appendix~\ref{app:style-factors}.

\begin{table}[hbt]
	\caption{The 11 style factors in Exabel's factor model, categorized into the 8 factor groups proposed by MSCI.}
	\centering
	\begin{tabular}{ll}
		\toprule
		Factor group & Factors \\
		\midrule
        Volatility & Beta, Volatility \\
        Yield      & Dividend yield \\
        Quality    & Profitability \\
        Momentum   & Long and short term momentum \\
        Value      & Book value/price, Earnings yield \\
        Size       & Size \\
        Growth     & Sales growth \\
        Liquidity  & Liquidity \\
		\bottomrule
	\end{tabular}
	\label{tab:style-factors}
\end{table}


In order to reduce collinearity with $f_m$ and to enable meaningful interpretation, the style factor loadings are constrained to be market cap-centered, i.e.

\begin{equation*}
 \sum_{k\in\mathbb{K}_{\mathrm{eff}}} {mc}_k X_k^s = 0, \quad \mbox{ for all } s\in\mathbb{S},
\end{equation*}
where ${mc}_k$ is the market capitalisation of company $k$. If the style loadings are also standardized, the corresponding factor returns allow one to quantify to what degree a company's return on that day can be attributed to its exposure to a particular factor.

\subsection*{Country factors}

The {\em country} loadings represent a company's affiliation with one or several countries. If a company is unequivocally affiliated with one country $c$, then $X_k^c$ is equal to $1$ for this country and $0$ for all others. If a company is affiliated with multiple countries, this is indicated by values between $0$ and $1$ which sum up to $1$.
Fig.~\ref{fig:country-statistics} shows all countries represented in the effective estimation universe $\mathbb{K}_{\mathrm{eff}}$ on 1.1.2021 (i.e.\ all companies in $\mathbb{K}$ for which all factor loading data is available at this date) and depicts the number of companies from each country, as well as each country's share of the total market capitalization.

\subsection*{Industry factors}

The {\em industry} loadings represent a company's affiliation with one or several industries. If a company is unequivocally affiliated with one industry $i$, then $X_k^i$ is equal to $1$ for this industry and $0$ for all others. If a company is affiliated with multiple industries, this is indicated by values between $0$ and $1$ which sum up to $1$.

\begin{table}[hbt]
	\caption{The 14 different types of industries considered in Exabel's factor model.}
	\centering
	\begin{tabular}{ll}
		\toprule
        Business Services & Consumer Services \\
        Consumer Cyclicals & Energy \\
        Finance & Healthcare \\
        Industrials & Non-Energy Materials \\
        Consumer Non-Cyclicals & Technology \\
        Telecommunications & Utilities \\
        Other & Non-Corporate \\
		\bottomrule
	\end{tabular}
	\label{tab:industries}
\end{table}

\subsection*{Constraining the country and industry returns}

In order for all factor returns to be identifiable, both country and industry factor returns are constrained as follows
\begin{equation*}
 \sum_{c\in\mathbb{C}} {mc}_c f_c = 0, \quad \sum_{i\in\mathbb{I}} {mc}_i f_i = 0,
\end{equation*}
where ${mc}_c = \sum_{k\in\mathbb{K}_{\mathrm{eff}}} {mc}_k X_k^c$ is the total market cap for country $c$ and likewise for ${mc}_i$. 
Together with the constraints imposed on the style factors, these constraints allow an interpretation of $f_m$ as the market cap-weighted average return of the market, while the factor returns $f_c, c\in\mathbb{C}, f_i$, and $i\in\mathbb{I}$ quantify deviations of a company's return from the market return $f_m$ explained by its country and industry membership.


\section{Variance explained by the factor model}

In order for the different factors to be useful for modelling the co-movement of stock returns, they must be able to explain a considerable fraction of the variability of returns across the estimation universe. A common measure to quantify how well a model can explain the variability seen in the response variable is the {\em coefficient of determination}, here defined as
\begin{equation}
 R^2 = 1 - \frac{SS_{\mathrm{res}}}{SS_{\mathrm{tot}}} \label{eq:R2-definition}
\end{equation}
where $SS_{\mathrm{res}}$ is the residual sum of squares of the factor model, and $SS_{\mathrm{tot}}$ is the sum over the squared returns of all companies to which this model was fitted.

Since separate models are fitted for each date, we would normally obtain an $R^2$ value for each date and would have to summarize all of these values across time. However, we feel that it is especially important that the model performs well on days with large stock price movements, and this information gets lost when daily $R^2$ values are calculated. Therefore, we first aggregate the denominator and the enumerator in \eqref{eq:R2-definition} separately across time such that $SS_{\mathrm{res}}$ is now the residual sum of squares both across all companies and across the evaluation period, and likewise for $SS_{\mathrm{tot}}$. Then, we calculate a single $R^2$ value via \eqref{eq:R2-definition}. Days with large stock price movements thus have an increased influence on the overall sums of squares, while the larger magnitudes of the respective daily sums would cancel each other out. 

A market cap-weighted portfolio is a natural baseline for any factor investing strategy, and given the large number of (often small) companies in $\mathbb{K}$, it makes sense to emphasize a good fit of the factor model to companies with a larger market cap. In the cross sectional regression model \eqref{eq:factor-model} this is achieved by assuming the variance of the residual term $\varepsilon_k$ to be proportional to the inverse market cap, i.e.\ $\sigma_k^2=\sigma^2/{mc}_k$. This results in a weighted least squares fit to the factor loading data with weights proportional to each company's market cap. Consequently, when we calculate $SS_{\mathrm{res}}$ and $SS_{\mathrm{tot}}$ in \eqref{eq:R2-definition}, the calculation is also performed with (square root) market cap-weighted returns.
Four different $R^2$-based metrics are calculated:
\begin{enumerate}
 \item Both 1-day and 90-day\footnote{here defined as the sum of 1-day returns over a 90-day period} returns are evaluated
 \item Both in-sample and out-of-sample evaluation is performed. The former fits the factor model to all companies in $\mathbb{K}_{\mathrm{eff}}$ while the latter performs a 10-fold cross validation where $\mathbb{K}_{\mathrm{eff}}$ is split into 10 folds of approximately equal size. The factor model is fitted to 9 of the folds and evaluated on the remaining one, and this process is repeated such that each fold is left out once for evaluation.
\end{enumerate}

First, to get an idea about the relative importance of the market factor and the style, country, and industry factors, respectively, we show results for a reduced factor model based on just a subset of these factors:

\begin{table}[hbt]
 \caption{$R^2$ values for reduced factor models which use only the market factor, or, the market factor and either all style, all country, or all industry factors.}
 \centering
 \begin{tabular}{lllll}
  \toprule
  \multicolumn{1}{l}{}  & \multicolumn{2}{c}{in-sample} & \multicolumn{2}{c}{cross-validated} \\
 \midrule
                     &  1-day   &  90-day  &    1-day    &    90-day  \\
 \midrule
  {\it Market only}  &   0.171  &   0.227  &     0.169   &     0.225  \\
  Market + Style     &   0.280  &   0.339  &     0.264   &     0.326  \\
  Market + Country   &   0.339  &   0.335  &     0.318   &     0.314  \\  
  Market + Industry  &   0.215  &   0.295  &     0.200   &     0.282  \\  
  \bottomrule
 \end{tabular}
 \label{tab:R2-factor-groups}
\end{table}

Information about the companies' country membership seems particularly useful to explain the variability of their returns across $\mathbb{K}_{\mathrm{eff}}$. The style factors are equally important for explaining the 90-day returns but slightly less important for the 1-day returns. Information about a companies' industry membership also succeeds in explaining variability beyond the overall market movement, especially for the 90-day returns.

Next, we take a closer look at the importance of the individual style factors. As a reference, we now use a baseline model which uses the market factor and all country and industry factors, and we investigate how much additional variability can be explained by adding one style factor at a time:

\begin{table}[hbt]
 \caption{$R^2$ values for reduced factor models which use the market factor, all country and industry factors, and one additional style factor at a time.}
 \centering
 \begin{tabular}{lllll}
  \toprule
  \multicolumn{1}{l}{}  & \multicolumn{2}{c}{in-sample} & \multicolumn{2}{c}{cross-validated} \\
 \midrule
                         &  1-day   &  90-day  &    1-day    &    90-day  \\
 \midrule
  {\it no style factors} &   0.377  &   0.395  &     0.347   &     0.367  \\
  Beta                   &   0.389  &   0.415  &     0.358   &     0.388  \\
  Book value/price       &   0.381  &   0.403  &     0.350   &     0.375  \\  
  Dividend yield         &   0.382  &   0.401  &     0.350   &     0.373  \\  
  Earnings yield         &   0.380  &   0.401  &     0.349   &     0.372  \\
  Sales growth           &   0.380  &   0.399  &     0.349   &     0.371  \\  
  Liquidity              &   0.381  &   0.402  &     0.349   &     0.371  \\
  Short term moment.     &   0.383  &   0.399  &     0.352   &     0.371  \\
  Long term moment.      &   0.387  &   0.405  &     0.356   &     0.376  \\
  Profitability          &   0.380  &   0.401  &     0.348   &     0.372  \\
  Volatility             &   0.387  &   0.410  &     0.356   &     0.382  \\
  Size                   &   0.381  &   0.399  &     0.349   &     0.370  \\
  \bottomrule
 \end{tabular}
 \label{tab:R2-style}
\end{table}

Beta, volatility and (to a slightly lesser degree) long term momentum stand out as style factor that perform particularly well across all four metrics when it comes to explaining variability of returns across $\mathbb{K}_{\mathrm{eff}}$. All style factors add additional information relative to the baseline. That information, however, is not independent across the different factors because the underlying financial characteristics are correlated. We therefore show another set of results where now the baseline is the full factor model with all style, country, and industry factors, and the effect of removing one style factor at a time is investigated:

\begin{table}[hbt]
 \caption{$R^2$ values for the full factor model and reduced factor models in which one style factor at a time is omitted.}
 \centering
 \begin{tabular}{lllll}
  \toprule
  \multicolumn{1}{l}{}  & \multicolumn{2}{c}{in-sample} & \multicolumn{2}{c}{cross-validated} \\
 \midrule
                          &  1-day   &  90-day  &    1-day    &    90-day  \\
 \midrule
  {\it all factors incl.} &   0.415  &   0.447  &     0.376   &     0.412  \\
  -Beta                   &   0.410  &   0.438  &     0.371   &     0.403  \\
  -Book value/price       &   0.414  &   0.445  &     0.375   &     0.411  \\  
  -Dividend yield         &   0.414  &   0.446  &     0.375   &     0.411  \\  
  -Earnings yield         &   0.414  &   0.445  &     0.375   &     0.411  \\
  -Sales growth           &   0.414  &   0.445  &     0.375   &     0.411  \\  
  -Liquidity              &   0.414  &   0.445  &     0.376   &     0.412  \\
  -Short term moment.     &   0.412  &   0.445  &     0.374   &     0.411  \\
  -Long term moment.      &   0.411  &   0.443  &     0.372   &     0.409  \\
  -Profitability          &   0.414  &   0.445  &     0.376   &     0.411  \\
  -Volatility             &   0.412  &   0.444  &     0.374   &     0.410  \\
  -Size                   &   0.413  &   0.445  &     0.374   &     0.411  \\
  \bottomrule
 \end{tabular}
 \label{tab:R2-lofo}
\end{table}

Comparison with Table~\ref{tab:R2-style} shows that the contribution of each style factor on top of all remaining factors is substantially reduced compared to the scenario where just a single style factor is included in the factor model. Beta and long term momentum provide the largest benefit, while other factors like liquidity provide little or no additional information that is quantifiable within the 3-digit rounding precision used in this table. The latter were still included in the factor model since all of the style factors on our list (Table~\ref{tab:style-factors}) have a distinct interpretation, are widely used in the context of factor investing, and have a strong theoretical foundation behind them. The results in Table~\ref{tab:R2-lofo} show, however, that it becomes increasingly difficult to add factors which provide truly independent information. Therefore, and to maintain good interpretability, we have refrained from adding additional style factors to our list. Several alternatives to the ones listed in Table~\ref{tab:style-factors} were considered, the associated $R^2$ values and our rationales for not including them are provided in a separate document.


\section{Examples of factor-based portfolios}

\begin{figure*}[ht]\centering 
	\includegraphics[width=\linewidth]{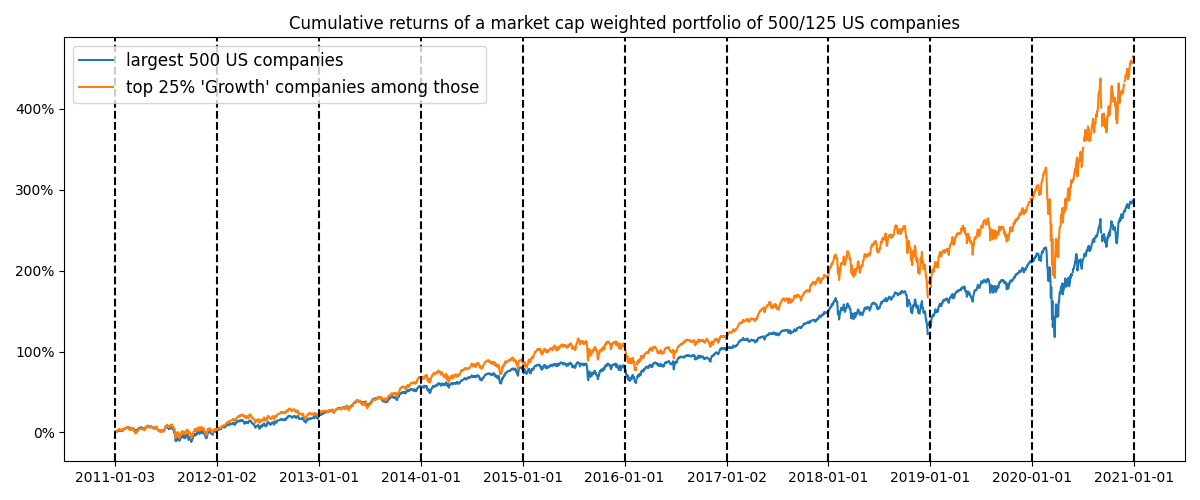}
	\caption{Cumulative returns of a portfolio consisting of the largest 500 US companies and a portfolio consisting of the subset of the 25\% companies with the largest 'Growth' factor loading.}
	\label{fig:growth-portfolio-returns}
\end{figure*}

To illustrate how creating a portfolio with increased exposure to a particular factor can affect that portfolio's performance, we compare the returns of a portfolio consisting of the largest (by market cap) 500 US companies, weighted proportionally to their market cap, with portfolios constructed as
\begin{description}[noitemsep]
 \item[Value:] a subset of the 125 companies with the largest 'Book value/price' and 'Earnings yield' factor loadings\footnote{ranks are first calculated separately for the two factors, then a ranking of the average ranks of those two components is used to select the top 125 'Value' companies. The same procedure is used for the 'Momentum' portfolio.}
 \item[Low size:] a subset of the 125 companies with the smallest size (by market cap)
 \item[Momentum:] a subset of the 125 companies with the largest 'Short term momentum' and 'Long term momentum' factor loadings
 \item[Quality:] a subset of the 125 companies with the largest 'Profitability' factor loadings
 \item[Yield:] a subset of the 125 companies with the largest 'Dividend Yield' factor loadings 
 \item[Low volatility:] a subset of the 125 companies with the lowest 'Volatility' factor loadings 
 \item[Growth:] a subset of the 125 companies with the largest 'Sales growth' factor loadings
 \item[Liquidity:] a subset of the 125 companies with the largest 'Liquidity' factor loadings 
\end{description}
The 125 companies within those subsets are again weighted proportionally to their market cap. The weights of the market cap portfolio and the factor-based portfolios are adjusted at the beginning of each month according to the updated figures for market cap and factor loadings.

Fig.~\ref{fig:growth-portfolio-returns} depicts the cumulative returns of the 'Growth' portfolio and shows that during the 2011 - 2020 period this portfolio would have outperformed the market cap portfolio with the 500 largest companies. The differences are especially pronounced during the period of the stock market recovery after the Coronavirus crash of 2020. The portfolios depicted in Fig.~\ref{fig:growth-portfolio-returns} mimic those that would be obtained with portfolios based on the S\&P 500 and the S\&P 500 Pure Growth index, respectively, and so the patterns of the cumulative return curves are very similar to those presented in  \cite{SP500-PureGrowth}.

\begin{figure*}[ht]\centering
	\includegraphics[width=14cm]{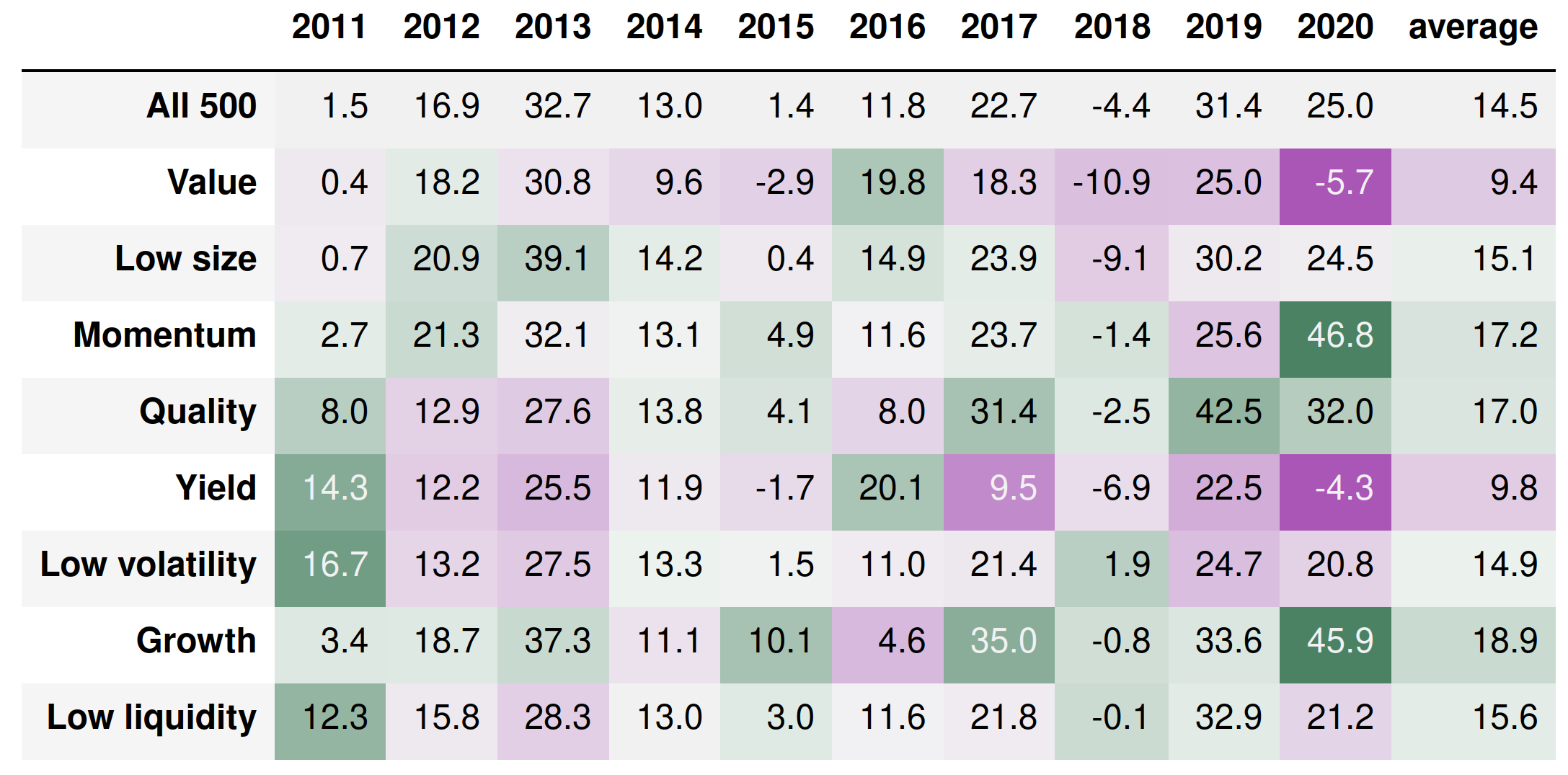}
	\caption{Annual returns (in \%) of portfolios that emphasize different factors. Years in which a factor based portfolio outperformed the market cap portfolio are highlighted in green, years of underperformance are highlighted in purple. Darker colors indicate more pronounced differences in performance.}
	\label{fig:portfolio-returns-table}
\end{figure*}

The relative performance of the factor-based portfolios varies during the economic cycle. The table in Fig.~\ref{fig:portfolio-returns-table} shows annual returns for each calendar year from 2011 to 2020 for the factor-based portfolios described above. The average performance over the 10-year period is also shown and calculated as
\begin{equation*}
 r_{\mathrm{10y}} = \left(\prod_{y=2011}^{2020} (1+r_y)\right)^{\frac{1}{10}} -1
\end{equation*}
where $r_y$ denotes the annual return in year $y$. We note that while there is no guarantee that a particular factor exposure is beneficial over a given time period, the results in Fig.~\ref{fig:portfolio-returns-table} are in agreement with the asset pricing literature in that over extended time periods, many factors are able to earn a premium. In any case, awareness of the their portfolios' exposures can help investors make more informed investment decisions.

\appendix

\section{Style factor definitions}
\label{app:style-factors}

We provide a brief explanation and definition of the 11 {\em style} factors used in Exabel's factor model. Since the factor loadings are defined individually for each company, we drop the subscript $k$, which was used above to indicate the dependence on the company. Several factor loadings are calculated from data during a certain time period $\mathbb{T}$, so we add a subscript $t$ to indicate dependence on time, e.g.\ we now write $r_t$ for the return on date $t$. We denote by $t_0$ the date index at which the factor loading is (re-)calculated.

All factor loadings are updated monthly, based on the most recent available data. Data which is only available quarterly or semi-annually is forward filled for up to six months. The values from the calculations detailed below are normalized by a robust scaler which maps the 25th percentile to $-1$ and the 75th percentile to $1$. Subsequently, the normalized values are clipped at $\pm3$ and market cap-centered.

\subsection*{Beta}
This factor measures to what degree the movements in the price of a particular stock are explained by movements in the market. We calculate it by regressing the (overlapping) 7-day returns $r_{t,7d}$ of each company during the last 364 days on the 7-day return of the NASDAQ Global Index $\bar{r}_{t,7d}$ over the same time period:
\begin{equation*}
 r_{t,7d} = \alpha + \beta \bar{r}_{t,7d} + \epsilon_t
\end{equation*}
The intercept parameter $\alpha$ is not used any further, while the slope parameter $\beta$ defines our factor loading. A large value indicates that a company's return is strongly correlated with the overall market return, whereas a small value suggests that the movements of this company's stock price are largely independent from the market.

\subsection*{Volatility}
This factor measures the magnitude of the fluctuations in the stock price movements. We calculate it as the standard deviation of daily returns over the last 91 days:
\begin{equation*}
 \mathrm{DStD} = \sqrt{\frac{1}{91}\sum_{t\in\mathbb{T}}\left(r_t-\frac{1}{91}\sum_{t\in\mathbb{T}} r_t\right)^2},
\end{equation*}
where $\mathbb{T}:=\{t_0-91, \ldots, t_0-1\}$.
A stock with large volatility is a riskier investment with a higher probability for both large gains and large losses.

\subsection*{Dividend yield}
This factor loading is calculated as the ratio of dividends (over the last year) per share to the share price: 
\begin{equation*}
 \mathrm{DtP} = \frac{\mathrm{dividends~per~share}}{\mathrm{price~per~share}}.
\end{equation*}
A larger dividend yield is often associated with mature companies while small, faster growing companies tend to pay a lower average dividend. It should also be noted that falling share prices entail a higher dividend yield if the dividend is kept constant.

\subsection*{Profitability}
A company's profitability is here measured as the `return on average assets', i.e.\ the ratio of its net income to total assets:
\begin{equation*}
 \mathrm{ROA} = \frac{\mathrm{net~income}}{\mathrm{total~assets}}.
\end{equation*}
This `quality' factor thus quantifies how efficiently a company uses its assets to generate earnings.

\subsection*{Momentum}
Momentum factors quantify trends in the stock's returns. In Exabel's model two different momentum factors are included which capture short term (`$\mathrm{STM}$') and longer term (`$\mathrm{LTM}$') trends, respectively. If we denote the share price at time $t$ by $P_t$, the two momentum loadings are defined as 
\begin{equation*}
 \mathrm{STM} = \frac{P_{t_0-1}-P_{t_0-29}}{P_{t_0-29}}, \qquad \mathrm{LTM} = \frac{P_{t_0-29}-P_{t_0-365}}{P_{t_0-365}} \; .
\end{equation*}

\subsection*{Book value/price}
This `value' factor measures a company's book value relative to its market valuation (share price), i.e.
\begin{equation*}
 \mathrm{BtP} = \frac{\mathrm{book~value}}{\mathrm{price~per~share}}.
\end{equation*}

\subsection*{Earnings yield}
Another `value' factor, this one measures a company's earnings per share (EpS) relative to its market valuation, i.e.
\begin{equation*}
 \mathrm{EtP} = \frac{\mathrm{earnings~per~share}}{\mathrm{price~per~share}}.
\end{equation*}
When available, the analysts' consensus estimate of the EpS for the next year is considered in addition to the company's reported EpS for the prior 12 months, and it is given three times the weight of the reported EpS. If such estimate is not available, only the reported EpS is used for the calculation of the factor loadings.

\subsection*{Size}
This factor quantifies the company's size, given here as the logarithm of its market capitalization (mc):
\begin{equation*}
 \mathrm{LgMC} = \log_{10}(\mathrm{mc}).
\end{equation*}
Its relevance is based on the finding that companies with smaller market cap tend to have higher risk adjusted returns during certain periods of the economic cycle \cite{Banz:1981}.

\subsection*{Sales growth}
This factor aims to capture a company's growth prospects. We calculate it in the same way as described by S\&P in the construction of their S\&P 500 Pure Growth index \cite{SP500-PureGrowth}: 
\begin{itemize}[noitemsep]
 \item when sales data $S_t$ from three years prior are available
 \begin{equation*}
  \mathrm{SGr} = \frac{S_{t_0-1}-S_{t_0-1095}}{S_{t_0-1095}} \; / \; 3
 \end{equation*}
  \item when sales data $S_t$ from two years prior are available
 \begin{equation*}
  \mathrm{SGr} = \frac{S_{t_0-1}-S_{t_0-730}}{S_{t_0-730}} \; / \; 2
 \end{equation*}
  \item when sales data $S_t$ from only year prior are available
 \begin{equation*}
  \mathrm{SGr} = \frac{S_{t_0-1}-S_{t_0-365}}{S_{t_0-365}}
 \end{equation*}
\end{itemize}

\subsection*{Liquidity}
A company's liquidity is here quantified as the logarithm of the average daily share turnover over the last 91 days
\begin{equation*}
 \mathrm{STO}_\mathrm{91d} = \log\left(\frac{1}{91}\sum_{t\in\mathbb{T}}\frac{V_t}{SO_t}\right),
\end{equation*}
where $\mathbb{T}:=\{t_0-91, \ldots, t_0-1\}$, $V_t$ is the trading volume and $SO_t$ is the number of shares outstanding on date $t$. The latter is the aggregate number of shares that a corporation has issued to investors.


\phantomsection
\bibliographystyle{unsrt}
\bibliography{bibliography.bib}

\begin{thebibliography}{1}

\bibitem{MSCI-FaCS}
{MSCI}.
\newblock {FaCS: Introducing our latest Factor Innovation – MSCI
  FaCS\textsuperscript{TM}}.
\newblock \url{https://www.msci.com/our-solutions/factor-investing/facs}.
\newblock [Online; accessed 1-November-2021].

\bibitem{SP500-PureGrowth}
{S\&P Dow Jones Indices}.
\newblock {S\&P 500 Pure Growth}.
\newblock
  \url{https://www.spglobal.com/spdji/en/indices/equity/sp-500-pure-growth/}.
\newblock [Online; accessed 1-November-2021].

\bibitem{Banz:1981}
R.~W. Banz.
\newblock The relationship between return and market value of common stocks.
\newblock {\em Journal of Financial Economics}, 9(1):3--18, 1981.

\end{thebibliography}


\end{document}